\documentclass[prd,amsmath,superscriptaddress,nofootinbib,showpacs,twocolumn]{revtex4}
\usepackage{bm}
\usepackage{amsmath}
\usepackage{graphicx}
\usepackage{color}
\def\bfm#1{\mbox{\boldmath $#1$}}

\newcommand{\ie}{{i.e.}}
\begin{document}
\title{
The fate of pion condensation in quark matter: \\
from the chiral limit to the physical pion mass }
\author{H. Abuki}
\email{abuki@th.physik.uni-frankfurt.de}
\affiliation{Institut f\"ur Theoretische Physik, J.W. Goethe
Universit\"at, D-60438 Frankfurt am Main, Germany}
\author{R. Anglani}\email{roberto.anglani@ba.infn.it}
\affiliation{Istituto Nazionale di Fisica Nucleare (INFN), Sezione di
Bari, I-70126 Bari, Italy}
\affiliation{Dipartimento di Fisica, Universit\`a di Bari, I-70126 Bari,
Italy}
\author{R. Gatto}
\email{raoul.gatto@physics.unige.ch}
\affiliation{D\'epartement de Physique Th\'eorique, Universit\'e de
Gen\`eve, CH-1211 Gen\`eve 4, Switzerland}
\author{M. Pellicoro}
\email{mario.pellicoro@ba.infn.it}
\affiliation{Istituto Nazionale di Fisica Nucleare (INFN), Sezione di
Bari, I-70126 Bari, Italy}
\affiliation{Dipartimento di Fisica, Universit\`a di Bari, I-70126 Bari,
Italy}
\author{M. Ruggieri}
\email{marco.ruggieri@ba.infn.it}
\affiliation{Istituto Nazionale di Fisica Nucleare (INFN), Sezione di
Bari, I-70126 Bari, Italy}
\affiliation{Dipartimento di Fisica, Universit\`a di Bari, I-70126 Bari,
Italia}

\preprint{BA-TH/602-09}

\begin{abstract}
We study aspects of the pion condensation in two-flavor neutral
  quark matter using the Nambu--Jona-Lasinio model of QCD at
  finite density.
 We investigate the role of electric charge neutrality,
  and explicit symmetry breaking via quark mass, both of
  which control the onset of the charged pion $(\pi^c)$ condensation.
 We show that the equality between the electric chemical potential and
  the in-medium pion mass, $\mu_{e}=M_{\pi^-}$, as a threshold, persists
  even for a composite pion system in the medium, provided the transition to
  the pion condensed phase is of the second order.
 Moreover, we find that the pion condensate in neutral quark matter
  is extremely fragile to the symmetry breaking effect via a current
  quark mass $m$, and is ruled out for $m$ larger than the order of 10
  keV.
\end{abstract}

\maketitle 
\section{Introduction}
In the early 1970s Migdal suggested the possibility of 
 pion condensation in a nuclear medium \cite{Migdal,picond}.
Since then, many efforts have been made in order to clarify the
 in-medium pion properties affected by the pion-nucleon interaction
 \cite{Kunihiro:1993pz}, because a better insight of such condensation
 phenomena would yield important advances not only in subnuclear physics
 such as that of pionic atoms \cite{Kienle:2004hq}, but also in the
 physics of neutron stars \cite{Maxwell:1977zz}, supernovas
 \cite{Ishizuka:2008gr},
 and the heavy ion collisions \cite{Zimanyi:1979ga}.
From the results of the experiments concerning the repulsive $\pi N$
 interaction, the simplest possibility, the $S$-wave pion
 condensation is highly unlikely to be realized in nature.
In contrast, the possibility of $P$-wave pion condensation has been
 widely argued \cite{picond,EricsonWeise,Wakasa:1997zz}, even though it
 still remains as a matter of debate if this occurs at several times the
 ordinary nuclear density.
Most of the studies about this issue performed so far are mainly concerned
 with the role of the coupling between pions and baryons in the nuclear
 medium \cite{Kienle:2004hq}, considering the pion itself as an
 elementary object.
However, since the pion is considered to be the Nambu-Goldstone boson
 associated with chiral symmetry breaking, its {\em internal structure}
 and {\em mass} may also be sensitive to the modification of the QCD
 vacuum itself in the finite density environment \cite{Hatsuda:1994pi};
the finite baryon density, and even the isospin density arising from the
 neutrality condition, can modify the structure of the QCD ground state, and
 it can in turn produce significant modifications of the pion properties
 in the medium.

In this article we revisit the possibility of $S$-wave charged pion
($\pi^c$) condensation starting from a {\em microscopic} model which is
 built with quarks as the constituents of pions, and which exhibits
 chiral symmetry restoration at the finite quark chemical 
 potential $\mu$ or temperature $T$.
This strategy also enables us to present a unified view on the possible
 crossover from a Bose-Einstein condensate (BEC) of $\pi^-$ condensate
 to a BCS-type $\langle\bar{u}\gamma_5d\rangle$ superfluid
 (BCS-BEC crossover) at finite isospin density \cite{He:2005nk}.
 This possibility was first speculated by applying the chiral Lagrangian at
 low $\mu_I$ and the perturbative QCD result at high $\mu_I$
 \cite{Son:2000by}; 
 since then, the QCD phase structure at finite isospin density has been
 widely investigated using  the chiral effective models
 \cite{Toublan:2003tt,Mao:2006zr,Andersen:2008qk},
 a lattice-based model \cite{Nishida:2003fb},
 the random matrix model \cite{Klein:2004hv},
 and the lattice QCD simulations
 \cite{Kogut:2002zg,deForcrand:2007uz,Detmold:2008yn}.

Since $\mu$ acts as the external field producing a mismatch in $u$-$\bar{d}$
 Fermi surfaces \cite{Son:2000by,He:2006tn},
 our problem also closely intersects with the
 problem of the BCS-BEC crossover with a density imbalance in the cold
 atomic systems \cite{coldatom}.
Finally, it should be noted that in
 Refs.~\cite{Ebert:2005wr,Andersen:2007qv} some window
 was found for the $S$-wave $\pi^c$ condensate on the finite $\mu$ axis
 in the Nambu--Jona-Lasinio (NJL) model.
This was called the ``{\em gapless pion condensate}'' since the
 constituent quark presents a gapless dispersion relation,
 due to $u$-$\bar{d}$ Fermi surface mismatch.
Our purpose here includes clarification of how this can be understood
 consistently with the number of negative results on the $S$-wave pion
 condensate obtained in the past.
To this end, we derive the appropriate criterion for $\pi^c$
 condensation, and based on this criterion, we show that the $\pi^c$
 condensation found in \cite{Ebert:2005wr} is extremely fragile to
 explicit chiral symmetry breaking due to a finite current quark mass.

The purpose of this article is twofold. First, we investigate the
 conditions for the onset of $\pi^c$ condensation at finite density
 using the NJL model of QCD; in this model pions are not elementary
 fields, but they are described as the bound states of quarks.
We find that even in such a composite pion model with dynamical quarks,
 the threshold for $\pi^c$ condensation for noninteracting elementary
 pion gas, $\mu_e = M_{\pi^-}$ ($-\mu_e=M_{\pi^+}$) for positive
 (negative) $\mu_e$ (with $\mu_e$ being the electric chemical
 potential), holds as well.
This enables us to elucidate the effect of the current quark mass in the
 neutral ground state, which will be summarized by drawing the phase
 diagram in the $(\mu,\,m_\pi)$ plane, where $m_\pi$ denotes the vacuum pion
 mass.
Furthermore we investigate the effect of $\mu_e$ at the physical value
 of the current quark mass drawing a phase diagram in the $(\mu,\,\mu_{e})$ 
 plane.
Based on these analyses, we conclude that $\pi^c$ condensation is
 forbidden in realistic neutral quark matter.

\section{The model}
We study two-flavor quark matter at a finite chemical potential within
 the NJL model. The Lagrangian of the model is given
 by~\cite{Abuki:2008tx}
\begin{eqnarray}
{\cal L} &=& \bar{e}(i\gamma_\mu\partial^\mu + \mu_e \gamma_0)e +
 \bar\psi\left(i\gamma_\mu \partial^\mu + \hat\mu\gamma_0 -m\right)\psi
 \nonumber \\
&&+ G\left[\left(\bar\psi \psi\right)^2 + \left(\bar\psi i \gamma_5
        \vec\tau \psi\right)^2\right]~, \label{eq:Lagr}
\end{eqnarray}
Here $e$ denotes the electron field, and $\psi$ is the quark spinor with
Dirac, color and flavor indices (implicitly summed). $m=m_u=m_d$ is the bare
 quark mass and $G$ is the coupling constant. $\mu_e$ is the electric
 charge chemical potential needed to keep the system
 electrically neutral \cite{Abuki:2008tx}, while $\mu_I$ serves as the
 isospin chemical potential in the hadron sector, $\mu_I=-\mu_e$
 since $Q=\frac{1}{2}B+I_3$.
The quark chemical potential matrix $\hat\mu$ is defined in flavor-color
 space as $\hat\mu=\text{diag}(\mu-\frac{2}{3}\mu_e,
 \mu+\frac{1}{3}\mu_e)\otimes\bm{1}_c$,
 where $\bm{1}_c$ denotes the identity matrix in color space, $\mu$ is
 the quark chemical potential related to the conserved baryon number.
By performing the mean field approximation, we examine the possibility
 that the ground state develops condensation in the $\sigma =
 G\langle\bar\psi\psi\rangle$
 and/or $\bfm{\pi}=G\langle\bar\psi i\gamma_5\bfm{\tau}\psi\rangle$
 channels, where $\bfm{\tau}=\{\tau_1,\tau_2,\tau_3\}$ denotes the Pauli
 matrices. \footnote{In this study we restrict ourselves to the
 homogeneous ground state, although some interesting possibilities of
 spatially modulated condensate are argued in Ref.~\cite{Nakano:2004cd}.}
We find that $\langle\pi_3\rangle$ is always zero, as there is no driving
 force, so we omit it in the following.
We use the notation, $M=m-2\sigma$ and $N=2\sqrt{\pi_1^2+\pi_2^2}$.
In the numerical analyses, we fix $\Lambda=651$ MeV and $G=2.12/\Lambda^2$
 so that the model reproduces $f_\pi=92$ MeV,
 $\langle\bar{u}{u}\rangle=-(250\,{\rm MeV})^3$,
 and $m_\pi=139$ MeV in the vacuum with $m=5.5$ MeV.

\section{The symmetry structure}
The global symmetry of the model at $m=\mu_e=0$ is
the $SU_{\rm L}(2)\times SU_{\rm R}(2)$ chiral symmetry.
This symmetry will be broken spontaneously to $SU(2)$ by the emergence
 of a nonzero condensate $\sqrt{\sigma^2+\bfm{\pi}^2}$.
When the finite quark mass $m$ is turned on, the original chiral
 symmetry is explicitly broken to the diagonal
 subgroup ${SU}_{\rm L+R}(2)$.
This forces the chiral condensate to the $\sigma$ direction
 (vacuum alignment), and the pions as Goldstone bosons acquire a mass,
 but they remain degenerate in the isospin triplet.
The effect of a small $\mu_I$ is then to split the masses of the three
 pions, since the diagonal $SU_{\rm L+R}(2)$ is broken explicitly
 to $U_{\tau_3}(1)$,
 which denotes the rotation about the isospin third axis.
When $\pi_1$ and/or $\pi_2$ condensates (or simply $N$) become nonzero,
 this $U_{\tau_3}(1)$ symmetry gets spontaneously broken
 and we expect one massless Goldstone boson to appear.
It is interesting to note that when the discrete parity is considered
 together, there remains one discrete symmetry even in the pion
 condensed phase. The parity operation acts
 as $\pi\to\mathcal{P}\pi=-\pi$.
On the other hand, $Z_2\subset U_{\tau_3}(1)$ defined
 by $\mathcal{U}\in Z_2 \Leftrightarrow\mathcal{U}=e^{i\frac{\pi}{2}\tau_3}$,
 also flips the sign of the $\pi^c$ condensate.
Hence, the combined transformation, a
 {\em rotated parity} $P'=P\cdot Z_2$,
 remains unbroken.
The new symmetry breaking pattern of the $\pi^c$ condensate can be
 specified as $U_{\tau_3}(1)\times P\stackrel{\pi\ne0}{\to} P'$.
The new parity $P'$ is analogous to what is discussed in
 \cite{Kryjevski:2004cw}.
Finally, $\mu$ breaks the charge conjugation, causing stress which 
 produces a mismatch in $u$ and $\bar{d}$ densities that are equal at
 small (large) $\mu$ ($\mu_I$).

\section{Pion condensation at finite $\mu_I$ and at finite $\mu$}
Before discussing $\pi^c$ condensation in the medium, it is
instructive to review the classical arguments at $\mu=0$~\cite{Son:2000by}.
The shift of the vacuum energy caused by the pion fields at the lowest
 nonvanishing order in the fields $\bm \pi$ is
 $ \delta\Omega=(m_\pi^2+\mu_e^2)\pi_0^2/4 + (m_\pi^2 -
 \mu_e^2)\pi_+ \pi_-/2\,$.
From this we can infer that in the chiral limit ($m_\pi = 0$) an
 infinitesimally small value of $\mu_I$ favors condensation,
 $\langle\pi_+\pi_-\rangle\ne0$.
In the general case of $m_\pi \neq 0$ and $\mu_e \neq 0$, the condition
 $|\mu_e| > m_\pi$
 needs to be fulfilled so that the $\pi^c$ condensation turns out to be
 energetically favored.
In this kind of approach, using chiral perturbation theory,
 the chiral $SU(2)$ multiplets are elementary fields whose internal
 structures do not suffer from any quantum corrections.
On the other hand, the same threshold at $\mu=0$
 is obtained also in the NJL-type model~\cite{He:2005nk}.

\begin{figure}[t]
\begin{center}
\includegraphics[width=8cm]{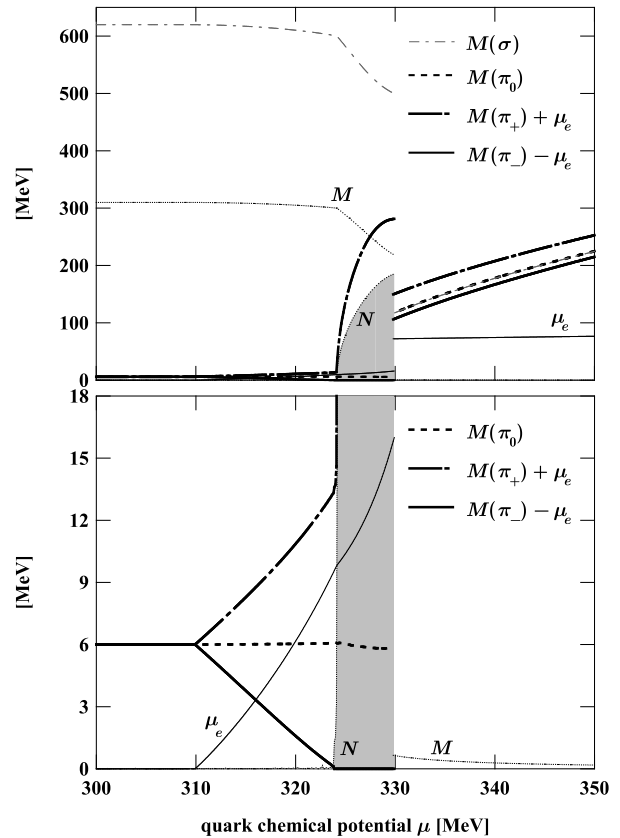}
\end{center}
\caption{
The constituent quark mass $M$, the pion
 condensate $N$, and meson masses as a function of $\mu$
 at $T=0$ in the neutral phase for a toy value of the current 
 quark mass $m=10$ keV.}
\label{fig:masse}
\end{figure}

We now consider neutral quark matter at $\mu\neq0$ and $T=0$. We are
 interested in the relation between the threshold of $\pi^c$
 condensation at finite density and the in-medium pion masses in the
 neutral ground state.
We have shown in Ref.~\cite{Abuki:2008tx} that at the physical
 point $m=5.5$ MeV, there is no room for $\pi^c$ condensation in the
 neutral phase.
Similar conclusion is also obtained in \cite{Andersen:2007qv}.
This picture changes if we lower the current quark mass.
We discuss this point in detail later.
For our present purpose it is enough to state that we need a current
 quark mass of the order of 10 keV.
In Fig.~\ref{fig:masse} we plot $M$ and $N$ in the neutral phase as a
 function of $\mu$ for $m=10$ keV.
In this figure $M_{\pi^0}$, $M_{\pi^\pm}$ denote the in-medium pion
 masses defined by the poles of the pion propagators in the rest frame,
 computed in the random phase approximation (RPA) to the Bethe-Salpeter
 (BS) equation.
For charged pion sector in the phase of a vanishing pion condensate
 $N=0$, 
 the pole condition is $\Gamma_{\pi^+\pi^-}(Q_0;\mu,\mu_e,T)=0$
 with
 $\Gamma_{\pi^+\pi^-}(Q_0;\mu,\mu_e,T)=\frac{m}{2GM}-N_cQ_0^2J(Q_0;\mu_u,\mu_d,T)$
 being the inverse $\pi^+$ propagator,
 $Q_0=\omega-\mu_e$, $N_{c}$
 being the number of colors, and $J$ being the polarization defined by
\begin{equation}
\begin{array}{rcl}
&&J(Q_0;\mu_u,\mu_d,T)=\int_M^{\sqrt{\Lambda^2+M^2}}dE\,
 \frac{\sqrt{E^2-M^2}}{2\pi^2}\frac{1}{E^2-Q_0^2/4}\Big[1\\
&&-\sum\limits_{t=\pm}\left(\frac{Q_0+2tE}{2Q_0}f_F(E-t\mu_u)+\frac{Q_0-2tE}{2Q_0}f_F(E-t\mu_d)\right)\Big]\,,
\end{array}
\end{equation}
where $f_{F}(x)$ is the Fermi distribution.
We note that the polarization function is not symmetric under
 $Q_0\to-Q_0$
 if there is a finite isospin density in the system,
 $\langle\psi^\dagger\tau_3\psi\rangle\ne 0$;
 this asymmetry is responsible for the mass splitting
 of $M_{\pi^+}$ and $M_{\pi^-}$.
The excitation gaps for $\pi^+$ and $\pi^-$ are accordingly given
 by $(M_{\pi^+}+\mu_e)$ and $(M_{\pi^{-}}-\mu_e)$, corresponding to
 positive and negative solutions to the BS equation in $\omega$.
From Fig.~\ref{fig:masse} we notice that the transition to the pion
 condensed phase is of second order and it occurs at the point where
 $M_{\pi^-} = \mu_e$.

The numerical results shown in Fig.~\ref{fig:masse} can be understood by
a closer inspection to the BS  equation and the gap equation for pion
condensate.
Although the argument below applies even to the $T\ne0$ case, we
 set $T=0$ for simplicity.
For small $N$, we expand the free energy of the ground state up to
 fourth order in the charged pion condensate $N=2\sqrt{2\pi^+ \pi^-}$,
\begin{equation*}
 \Omega=\Omega_0+\alpha(\mu,\mu_e)\frac{N^2}{2} +
 \beta(\mu,\mu_e)\frac{N^4}{4},
\end{equation*}
 where $\Omega_0$ is the free energy at $N=0$. It can be shown that the
 second order coefficient $\alpha$ has the same structure as the $\pi^+$
 propagator, with $Q_0$ replaced by $-\mu_e$, \ie,
 $\alpha(\mu,\mu_e)=\frac{1}{4}\Gamma_{\pi^+\pi^-}(-\mu_e;\mu,\mu_e)$.
Provided that the transition to the $\pi^c$ condensed phase is of second
 order, the critical condition is given by the linearized gap equation,
 $\alpha(\mu,\mu_e)=0$,
 that determines the line of the critical electric chemical
 potential $\mu_e^c(\mu)$ in the $(\mu,\mu_e)$ plane.
From the equality of the pion propagator and $\alpha$ noted above, one
 immediately obtains the threshold $M_{\pi^\mp}=\pm\mu_e$ for positive
 (negative) $\mu_e$.
This is because $\Gamma_{\pi^+\pi^-}(-\mu_e,\mu,\mu_e)=0$ is also
 satisfied on the critical line $\mu_e^{c}(\mu)$, and this solution
 corresponds to the negative (positive) root of the BS
 when $\mu_e>0$ ($\mu_e<0$).

The above argument holds either with or without the neutrality
condition. The neutrality condition determines the neutrality line
 $\mu_e^{\rm neut}(\mu)$
 in the $(\mu\,,\,\mu_e)$ plane.
At $m=10$ keV, $\mu_e^c(\mu)$ and $\mu_e^{\rm neut}(\mu)$ intersect at
 the point $(\mu,\mu_e)\cong (324,10)$ MeV. 
This fact is somehow implicit in Fig.~\ref{fig:masse} where the physical
 quantities only along $\mu_e^{\rm neut}(\mu)$ is sketched.
We see from the figure that, as soon as $\mu$ exceeds the value of the
 vacuum constituent quark mass $M=310\,\mbox{MeV}$, $\mu_e$ becomes
 finite due to the appearance of quark Fermi surfaces.
 Accordingly, the mass degeneracy in the pion sector gets lifted.

When the critical point
$\mu=324\,\mbox{MeV}$($\equiv\mu_{c1},\,\mbox{hereafter}$%
 \footnote{%
 The critical point $\mu_{c1}$ for the transition
 from the chiral symmetry broken phase to the pion condensed phase
 corresponds to the critical baryon density $n_B\sim0.1\,n_s$, with $n_s$
 being the nuclear saturation density $0.16\,\mbox{fm}^{-3}$.
 The electron fraction is $Y_e\equiv n_e/n_B\sim0.0006$. These values
 are unconventionally small, which is simply attributed to the fact that
 we are working near the unphysical chiral limit, $m=10\,\mbox{keV}$.})
 is reached, $\mu_e$ hits $M_{\pi^-}$ from below, resulting in the
 gapless excitation in the $\pi^-$ mode.
 Thereafter, $\pi^-$ mode in the $N=0$ phase changes into
 the exact Nambu-Goldstone boson associated with $U_{\tau_3}(1)$
 breaking; the
 excitation stays gapless due to the mixing of $(\sigma,\pi^+,\pi^-)$
 modes in the BS equation for $N\ne0$ \cite{He:2005nk}.

 We also notice that the increase
 of $\mu_e$ is somehow quenched as soon as the system experiences the
 transition to the pion condensed phase.
 This is reasonable because the $\pi^-$ condensate tries to accommodate
 the $d$ quark Fermi surface that helps cancel the positive charge arising
 from $u$ quarks; then less electrons are needed.
 The pion condensed phase continues up
 to $\mu=330\,\mbox{MeV}(\equiv\mu_{c2},\,\mbox{hereafter})$
 where the system encounters the first order phase transition into
 the almost chirally symmetric quark matter phase;
 the phase is characterized by $M\sim m$. Since we are working near the
 chiral limit $m=10\,\mbox{keV}$,
 the mass splitting in the sigma and neutral pion sector is tiny.
 In fact $M_{\pi_0}$ is only
 slightly larger than $M_{\sigma}$; the difference is less
 than $1\,\mbox{MeV}$ which is actually invisible in the figure.

A few remarks are in order with regards to the nature of pion
condensed phase obtained here:
(i)~Having a look at the fermionic spectrum, we can see that it is of
gapless type~\cite{Ebert:2005wr}.
(ii)~An investigation of the spectral functions in the bosonic sector leads
us to conclude that it is the BEC of the real $d\bar{u}$ bound
state ($\pi^-$) rather than the BCS-like pionic superfluid.
Let us first address point (ii).
It is enough for our present purpose to focus on the onset of the pion
condensed phase.
At the onset $\mu=\mu_{c1}$, the $\pi^-$
mode becomes gapless, which means the inverse $\pi^-$ propagator
is zero at $\omega=0$.
On the other hand, looking at the spectral density, the imaginary part
of the $\pi^-$ propagator, it is easy to see that the threshold for
the $(d\bar{u})$ continuum at rest $(\bfm{Q}=0)$ is located
at $\omega_{\mathrm{th}}=2M-\mu_e(\equiv(M-\mu-\mu_e/3)+(M+\mu-2\mu_e/3))$.
Then, $(\omega=0)$ in the $\pi^-$ propagator is realized as an isolated
bound state pole if $2M>\mu_e$,
while in the opposite case it is realized as a soft mode peak whose
width goes to zero only when its momentum $\bfm{Q}$ goes to zero.
Actually the former is realized here, so the system is the BEC; the
isospin chemical potential $|\mu_I|=\mu_e$ is too small to realize the
BCS-like superfluid state \cite{Nishida:2005ds,Sun:2007fc}.
Let us now turn to point (i).
The dispersion relations for quasiquarks in the pion condensed
phase are given by
\begin{equation}
\begin{array}{rcl}
 E_{\tilde{u}}(p)&=&\sqrt{(\sqrt{M^2+p^2}+\mu_e/2)^2 + N^2}%
 -(\mu - {\mu_e}/{6}),\\[1ex]
 E_{\tilde{d}}(p)&=&\sqrt{(\sqrt{M^2+p^2}-\mu_e/2)^2 + N^2}%
 -(\mu - {\mu_e}/{6}),
\end{array}
\end{equation}
where the subscript $\tilde{u}$ ($\tilde{d}$) represents the quasiquark
which has a large $u$-quark ($d$-quark) content at large $p$.
Since we have already noted that $M>\mu_e/2$ in the BEC, each quasiquark
energy has a minimum at $p=0$.
Then we see that both quasiquarks have a blocking region 
in $p$-space, defined by the condition $E_{\tilde{u}/\tilde{d}}(p)<0$;
the constituent $u$ and $d$ quarks are, accordingly, accumulated in those
regions.
From (i) and (ii), we may conclude that the system is the BEC of
the real $d\bar{u}$ bound state ($\pi^-$) formed in the charge neutral
background characterized by the Fermi seas of $u$, $d$ quarks, and
electrons.

\section{The role of the current quark mass}
In the past years, a large number of the calculations dealing with the
 issue of $\pi^c$ condensation have been performed in the chiral limit.
 This treatment is simple from the point of view of calculations but
 somehow unphysical.
In this section we investigate the role of the current quark mass in
 $\pi^c$ condensation. In this analysis we fix the cutoff $\Lambda$
 and the coupling $G$ to the values specified above and we treat $m$ as
 a free parameter. As a consequence, the pion mass at $\mu=T=0$, $m_\pi$
 in the following, is a free parameter as well.
\begin{figure}[t]
\begin{center}
\includegraphics[width=8cm]{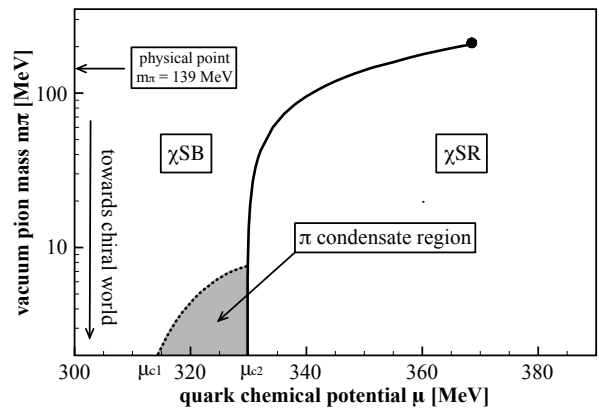}
\end{center}
\caption{
Phase diagram of neutral matter in $(\mu,\,m_{\pi})$ plane.}
\label{fig:phasempi}
\end{figure}
In Fig.~\ref{fig:phasempi} we report the phase diagram in
 $(\mu,\,m_\pi)$ plane in the neutral case.
The solid line represents the border between the two regions where 
chiral symmetry is broken and restored.
The bold dot is the critical endpoint of the first order transition.
The shaded region indicates the region where $\pi^c$ condensation
 occurs.
In the chiral limit $(m_{\pi}=0)$ our results are in good agreement with
 those obtained in Ref.~\cite{Ebert:2005wr}.
As we discussed in the previous section, there exist two
 critical values of the quark chemical potential, $\mu_{c1}$
 and $\mu_{c2}$, corresponding to the onset and vanishing of
 $\pi^c$ condensation, respectively.
Increasing the current quark mass results in the shrinking of the shaded
 region till the point $\mu_{c1}\equiv \mu_{c2}$
 for $m_{\pi}^{c}\sim 9\; \rm MeV$, corresponding to a current quark
 mass of $m \sim 10\;\rm keV$.
Hence we have shown that the gapless $\pi^c$ condensation is
 extremely fragile with respect to the symmetry breaking effect of 
 the current quark mass;
 this can be understood by observing the fact that the increase of quark
 mass leads to a drastic magnification of the vacuum pion mass at small $m$
 because of $m_\pi\propto\sqrt{m}$, while the change in quark mass in
 keV scale brings about no significant modification in $\mu_e$.

\begin{figure}[t]
\begin{center}
\includegraphics[width=8cm]{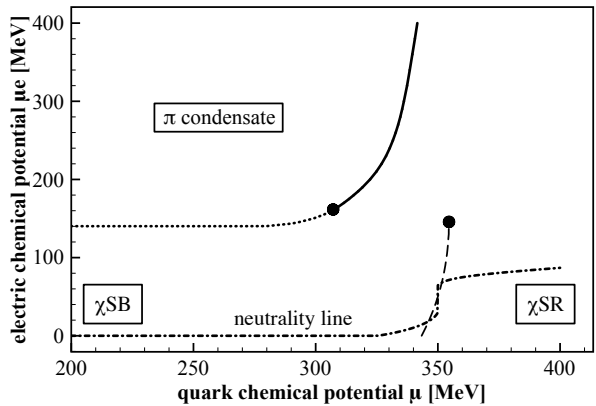}
\end{center}
\caption{
Phase diagram in $(\mu,\,\mu_{e})$ plane
 at $m=5.5$ MeV.}
\label{fig:phasediag} 
\end{figure}

\section{The phase diagram in the physical case}
As a final investigation, in Fig.~\ref{fig:phasediag} we draw the phase
 diagram of quark matter in the $(\mu,\,\mu_e)$ plane in the physical
 limit where the current quark mass is tuned to $m=5.5$ MeV.
At each value of $(\mu,\,\mu_e)$ we compute the chiral and pion
 condensates by minimization of the thermodynamical potential.
The solid line represents the first order transition from the $\pi^c$
 condensed phase to the chiral symmetry broken phase without the $\pi^c$
 condensate.
The bold dot is the critical endpoint for the first order transition,
 after which the second order transition sets in.
The dashed line indicates the first order transition between the two
 regions where chiral symmetry is broken and restored, respectively.
The dot-dashed line is the neutrality line
 $\mu_e^{\rm neut}=\mu_e(\mu)$
 which is obtained by requiring the global electrical neutrality condition,
 $\partial\Omega/\partial\mu_e = 0$.
The neutrality line manifestly shows the impossibility of finding a $\pi^c$
 condensate in a physical situation.

\section{Conclusions} 
In this article we have studied the phenomenology of $\pi^c$ condensation
 in two-flavor neutral quark matter using a Nambu--Jona-Lasinio model
 of QCD.
In particular, we have clarified the role of the quark mass $m$ and
the electric chemical potential $\mu_e$.
Our central results is that the threshold to the $\pi^c$ condensed phase,
 $\mu_e = M_{\pi^-}$,
 persists even for a composite pionic gas in the presence of dynamical
 quarks provided the transition to the $\pi^c$ condensed phase is of
 second order, which is indeed true up to $\mu\approx 300$ MeV.
Furthermore, we have found that the possibility of $\pi^c$ condensation
 is ruled out in neutral quark matter with a current quark mass
 larger than the order of 10 keV.
We have given a natural explanation for this fact by using the threshold
 criterion noted above.
As a further result we have computed the phase diagram in the
 $(\mu,\,\mu_e)$ plane at the physical quark mass. We find that the
 transition to the $\pi^c$
 condensed phase is of second order up to $\mu \approx 300$ MeV, after
 which it turns into a first order one.
By superimposing the neutrality line onto the phase diagram, we have
 shown that neutral quark matter never meets the regions of
 the nonvanishing $\pi^c$ condensate.
Thus we conclude that in neutral quark matter at the physical quark
 mass, the NJL model does not allow charged pion condensation
 in the ground state.

In the present analyses, we have focused on the modification of pion
propagation due to the character change of the QCD ground state with
respect to the finite baryon (and isospin) density environment. 
It should be noted, however, that in our restricted treatment, the
contribution to finite baryon density comes only
from constituent quarks and not from nucleons. 
The existence of nucleons, the bound states of constituent
quarks, should make $S$-wave $\pi^c$ condensation even less
favorable. This is because the Tomozawa-Weinberg interaction between
the nucleon and the pion is isospin odd, giving rise to repulsive pion 
self-energy at the physical situation where the neutron density is larger
than the proton density, \ie, $\langle n^\dagger n\rangle>\langle
p^\dagger p\rangle$ \cite{Muto:2003dk}. 
It would be interesting, though challenging, to investigate pion
condensation by taking into account nucleon degrees of freedom within the
NJL model. The prescription made in \cite{Bentz:2002um} may be useful.

\vspace*{1em}
{\em Acknowledgements.--} We thank D.~Blaschke, T.~Brauner,
P.~Colangelo, K.~Fukushima, D.~Rischke, A.~Sedrakian, A.~Ohnishi, and
T.~Tatsumi for a careful reading of the manuscript. The work of H.~A. was
supported by the Alexander von Humboldt Foundation. Part of numerical
calculations was performed using the facilities of the Frankfurt Center
for Scientific Computing.

\end{document}